\documentclass[conference]{IEEEtran}
\IEEEoverridecommandlockouts
\usepackage{cite}
\usepackage{amsmath,amssymb,amsfonts}
\usepackage{graphicx}
\usepackage{textcomp}
\usepackage{xcolor}

\def\BibTeX{{\rm B\kern-.05em{\sc i\kern-.025em b}\kern-.08em
    T\kern-.1667em\lower.7ex\hbox{E}\kern-.125emX}}
\begin{document}

\title{Event-Aligned Analysis of Multi-Rater Pain Assessments Using Continuous Wearable Physiology\\
}
\author{

\IEEEauthorblockN{Saba A. Farahani}
\IEEEauthorblockA{
University of California, Irvine\\
Irvine, CA, USA\\
fazizaba@uci.edu
}
\and

\IEEEauthorblockN{Elahe Khatibi}
\IEEEauthorblockA{
University of California, Irvine\\
Irvine, CA, USA\\
ekhatibi@uci.edu
}
\and

\IEEEauthorblockN{Thomas D. Hughes}
\IEEEauthorblockA{
University of California, Irvine\\
Irvine, CA, USA\\
thomasdh@hs.uci.edu
}
\and

\IEEEauthorblockN{Ariana M. Nelson}
\IEEEauthorblockA{
University of California, Irvine\\
Irvine, CA, USA\\
arianamn@hs.uci.edu
}
\and

\IEEEauthorblockN{Hung Cao}
\IEEEauthorblockA{
University of California, Irvine\\
Irvine, CA, USA\\
hungcao@uci.edu
}
\and

\IEEEauthorblockN{Amir M. Rahmani}
\IEEEauthorblockA{
University of California, Irvine\\
Irvine, CA, USA\\
a.rahmani@uci.edu
}
}

\maketitle
\begin{abstract}
Pain is assessed differently by patients, nurses, and clinicians, 
yet most computational approaches assume a single ground-truth 
label — effectively ignoring who is doing the rating. We introduce 
a rater-aware, event-aligned framework that converts sparse, 
rater-specific pain ratings into discrete pain-change events and 
aligns continuous wearable physiological signals to these events, 
preserving rater identity throughout. Applied to multimodal 
wearable data collected during spine-related pain procedures, the 
framework identifies substantial disagreement across rater groups 
and provides preliminary, exploratory evidence of rater-dependent 
physiological differences preceding reported pain increases. These 
findings suggest that pain-physiology relationships may not be 
rater-invariant, and that aggregating assessments across raters 
may mask meaningful physiological patterns. A rater-aware, 
event-aligned perspective is therefore a promising direction 
for interpreting wearable data in real-world clinical pain 
assessment.
\end{abstract}

\section{Introduction}
Pain assessment is essential in clinical care and biomedical 
research, informing diagnosis, treatment, and evaluation of 
outcomes. However, pain remains inherently subjective, shaped by 
individual perception, contextual factors, and observer 
interpretation. In clinical settings, multiple raters---including 
patients, nurses, and clinicians---assess pain using different 
information sources such as self-report, observable behaviors, and 
physiological indicators. As a result, pain assessments frequently 
differ among raters, even when collected at similar time points.

Recent advances in wearable sensing technologies enable 
continuous, noninvasive monitoring of physiological signals 
associated with autonomic and behavioral responses, including 
electrodermal activity (EDA), heart rate (HR), and skin 
temperature. These signals have shown associations with 
pain-related processes in both experimental and clinical 
settings~\cite{Chen2021,Cascella2023,Fang2025}. Despite this 
progress, integrating continuous physiological data with clinical 
pain assessments remains challenging: pain ratings are typically 
sparse, irregular, and rater-dependent, while physiological 
signals are dense and continuously evolving.

A key gap in current approaches is the assumption of a single 
ground-truth pain label, typically prioritizing patient self-report 
or aggregating ratings from multiple observers. Although this 
simplifies modeling, it may obscure meaningful differences in how 
pain is perceived and reported by different raters. Crucially, 
disagreement between raters should not be treated solely as 
annotation noise---it may instead reflect distinct and clinically 
meaningful assessment perspectives.

To address this gap, we present a rater-aware, event-aligned 
analysis framework that explicitly preserves rater-specific pain 
assessments. Rather than treating pain scores as static labels, 
the framework transforms sparse ratings into discrete pain-change 
events and aligns continuous wearable physiological signals to 
these events, maintaining temporal structure and enabling direct 
comparison of pre-event dynamics across rater groups.

We apply this framework to a clinically collected dataset 
comprising synchronized patient-, nurse-, and clinician-reported 
pain assessments alongside multimodal wearable signals recorded 
during spine-related pain procedures. This study makes three 
primary contributions:
\begin{enumerate}
    \item A rater-aware, event-aligned framework that converts 
    sparse pain ratings into rater-specific pain-change events;
    \item A systematic characterization of inter-rater disagreement 
    in pain-change reporting using clinically collected wearable 
    data; and
    \item Exploratory evidence that physiological dynamics preceding 
    reported pain changes differ across patients, nurses, and 
    clinicians, suggesting rater-dependent pain--physiology 
    relationships that warrant investigation in larger studies.
\end{enumerate}

\begin{figure}[t]
  \centering
  \includegraphics[width=\columnwidth]{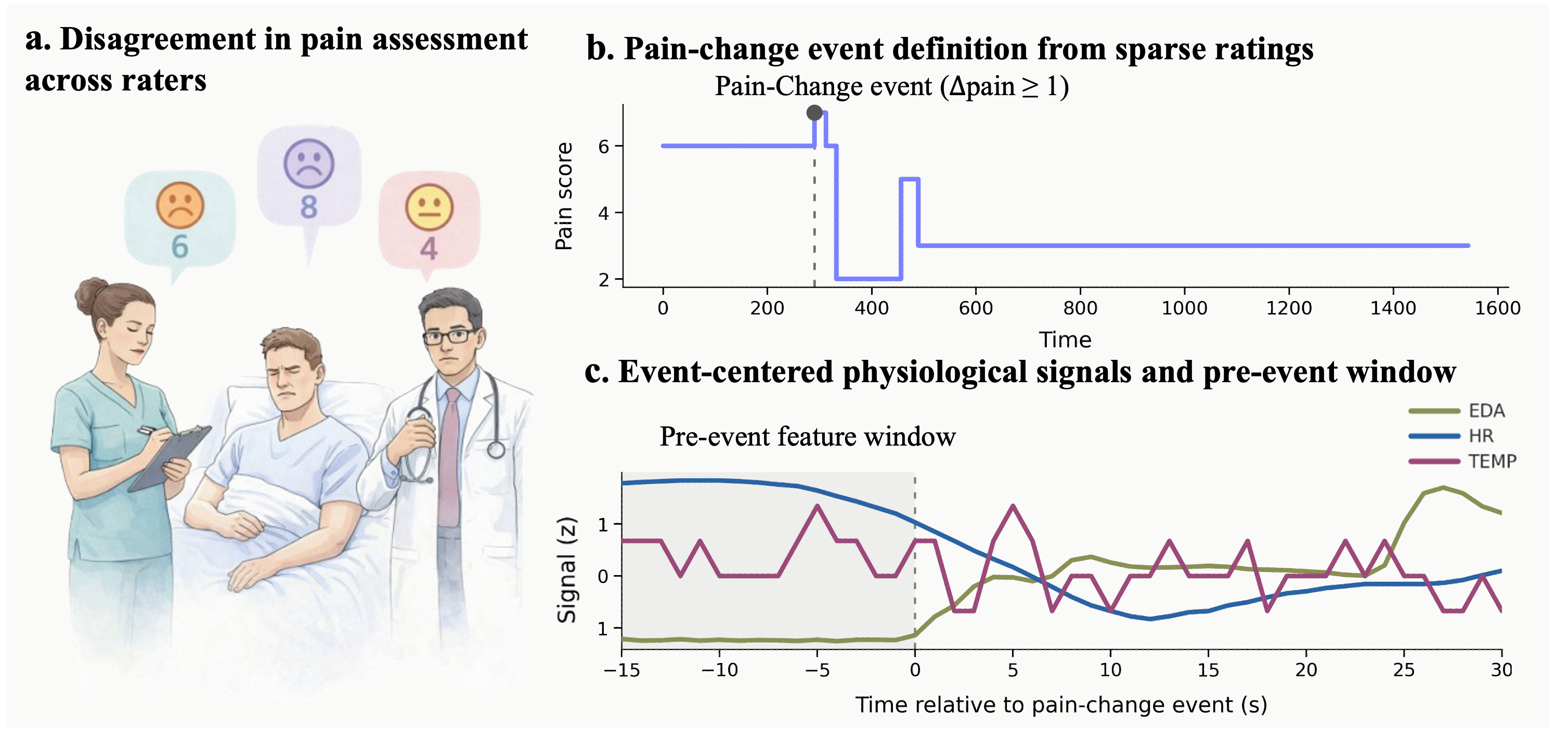}
  \caption{
  Overview of the proposed rater-aware, event-aligned analysis framework.
  (a) Pain assessments from patients, nurses, and clinicians may disagree even when collected at similar times.
  (b) Sparse pain ratings are converted into discrete pain-change events based on changes in reported pain scores.
  (c) Continuous wearable physiological signals are aligned to these events, and pre-event windows are extracted for rater-specific analysis.
  }
  \label{fig:overview}
\end{figure}

\section{Related Work}
Prior research has investigated automatic pain assessment using 
physiological signals and wearable sensors. These efforts have 
focused on autonomic and peripheral biomarkers, including 
electrodermal activity (EDA), heart rate and heart rate 
variability (HR/HRV), photoplethysmography (PPG), skin 
temperature, and motion, applying both classical machine learning 
and deep learning approaches~\cite{Chen2021,Cascella2023,Fang2025}. 
Laboratory and clinical evidence demonstrates that these signals 
reflect both nociceptive and affective dimensions of pain.

Multiple studies have established the feasibility of wearable-based 
pain assessment in clinical settings, particularly in postoperative 
and peri-procedural contexts. Models utilizing EDA and cardiovascular 
signals have achieved clinically meaningful accuracy in estimating 
pain intensity with wrist-worn devices 
\cite{Aqajari2021,Naeini2021,Subramanian2025}, and multimodal 
fusion techniques further enhance performance by integrating 
complementary physiological signals \cite{Thiam2020,Pouromran2021}. 
Nevertheless, most current methods represent pain as a single 
continuous variable and implicitly assume a unified ground truth 
among observers, which may obscure clinically meaningful 
differences in pain assessment.

Clinical and nursing literature, however, highlight that pain 
assessment is fundamentally subjective and rater-dependent. 
Clinical guidelines designate patient self-report as the reference 
standard when feasible, while recognizing that nurses and 
clinicians frequently rely on behavioral observations and 
physiological indicators when self-report is unavailable 
\cite{Devlin2018,Herr2019}. Empirical research consistently 
demonstrates substantial discrepancies between patient- and 
provider-reported pain, even when assessments occur at similar 
time points \cite{Hirsh2010,Rose2012}. Despite these findings, 
computational pain analysis seldom models rater disagreement 
explicitly, typically treating it as annotation noise rather than 
as a potentially informative signal.

A further challenge involves the temporal structure of pain data. 
Pain ratings are generally sparse and irregular, while 
physiological signals are collected continuously. Current 
approaches often address this mismatch using fixed windowing or 
label interpolation \cite{Yang2019,Subramaniam2021}, yet they 
typically treat pain scores as static labels rather than dynamic 
events reflecting fluctuations in perceived pain.

Context-conditioned graph models have likewise supported interpretable
analysis of dynamic neural physiology~\cite{Asadi2025DeepBrain,
Asadi2025BACE}. Although developed for deep-brain rather than wearable
signals, they motivate preserving temporal and behavioral context.

Taken together, these gaps motivate a rater-aware, event-aligned 
approach: one that preserves rater identity, models pain as a 
sequence of discrete change events, and aligns continuous 
wearable signals to those events for direct rater-specific 
comparison.

\begin{table}[h]
\caption{Study population and data characteristics.}
\label{tab:population}
\centering
\setlength{\tabcolsep}{3.5pt}
\renewcommand{\arraystretch}{1.15}
\begin{tabular}{|p{0.46\columnwidth}|p{0.48\columnwidth}|}
\hline
\textbf{Characteristic} & \textbf{Value} \\
\hline
Participants included in analysis & 24 \\
\hline
Age, mean $\pm$ SD (range)$^\dagger$ & 58.1 $\pm$ 11.3 (26--72) \\
\hline
Sex & 13 male / 10 female / 1 missing \\
\hline
Procedure type & Lumbar (15), Cervical (7), Other/unknown (2) \\
\hline
Wearable devices & Empatica E4 (n=19), EmbracePlus (n=5) \\
\hline
Physiological signals analyzed & EDA, HR, Skin temperature \\
\hline
Pain assessment raters & Patient, Nurse, Clinician \\
\hline
Pain scale & Numerical Rating Scale (0--10) \\
\hline
Pain assessments per subject & 3--6 per procedure \\
\hline
Analysis approach & Event-aligned pre-pain windows \\
\hline
\multicolumn{2}{|p{0.98\columnwidth}|}{\footnotesize $^\dagger$Range reflects one participant aged 26; all remaining participants were aged 42--72.} \\
\hline
\end{tabular}
\end{table}

\begin{figure*}[t]
  \centering
  \includegraphics[width=\textwidth]{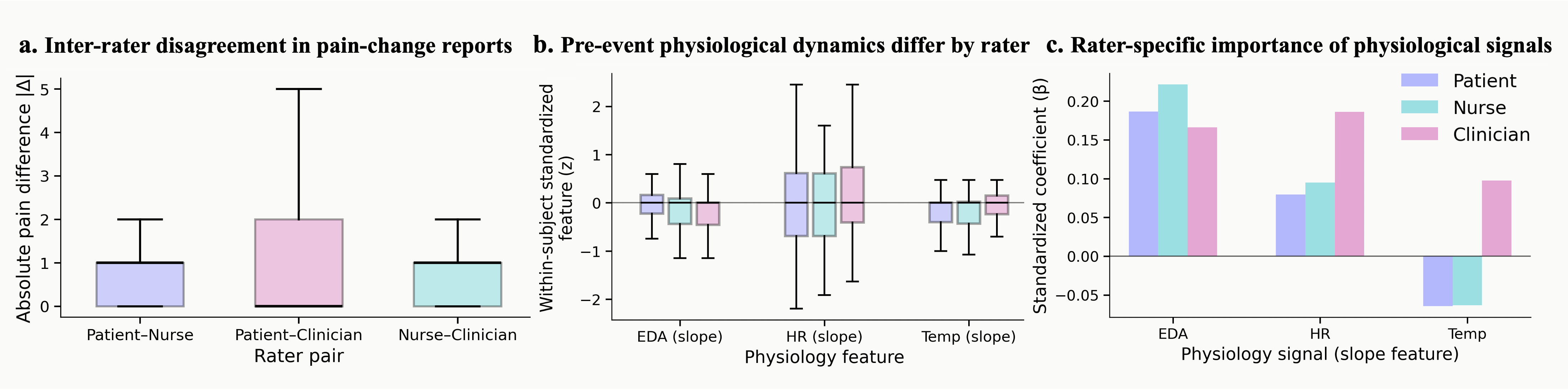}
  \caption{Summary of rater-dependent pain--physiology relationships.
  (a) Inter-rater disagreement in pain-change reports (absolute difference in reported pain change).
  (b) Distributions of pre-event physiological slope features by rater (within-subject standardized), highlighting rater-dependent dynamics.
  (c) Standardized linear-model coefficients fitted separately per rater, shown as an exploratory measure of relative signal association across physiological features; models are used for interpretability only and not for predictive evaluation.}
  \label{fig:results_overview}
\end{figure*}

\section{Method}

This study investigates differences in physiological dynamics 
preceding reported pain changes across various raters. To address 
this objective, a rater-aware, event-aligned analysis framework is 
presented. The framework transforms sparse pain ratings into 
discrete pain-change events and aligns continuous wearable 
physiological data to these events for comparison. 
Fig.~\ref{fig:overview} provides an overview of the proposed 
framework.

\subsection{Dataset and Study Setting}

A clinically collected dataset obtained during spine-related pain 
procedures was analyzed. This dataset comprises synchronized pain 
assessments from patients, nurses, and clinicians, along with 
continuous physiological signals recorded using wearable devices. 
Pain was reported at irregular intervals using the Numerical Rating 
Scale (NRS; 0--10).

Physiological data were collected continuously from wrist-worn 
wearable devices and include electrodermal activity (EDA), heart 
rate (HR), and skin temperature (Temp). These signals were selected 
based on their availability in wearable platforms and their 
established associations with autonomic and pain-related responses. 
Two wearable platforms were used: Empatica E4 (n=19) and 
EmbracePlus (n=5). Both devices record EDA at 4~Hz and derive HR 
from photoplethysmography at comparable effective rates; signals 
from both platforms were preprocessed identically. Device type was 
not systematically associated with rater group or procedure type 
in this cohort, though device effects cannot be fully excluded and 
represent a limitation acknowledged in Section~\ref{sec:discussion}. 
Participant demographics, procedure types, and dataset 
characteristics are summarized in Table~\ref{tab:population}. The 
experimental procedures involving human subjects described in this 
paper were approved by the Institutional Review Board of the 
University of California, Irvine.

\subsection{Rater-Specific Pain-Change Event Definition}

Pain ratings are typically sparse, irregular, and dependent on the 
individual rater, which complicates their alignment with continuous 
physiological signals. Instead of treating pain scores as static 
labels, changes in reported pain are modeled as discrete events.

For each rater independently, a pain-change event is defined at 
time $t$ when the absolute difference between two consecutive pain 
ratings exceeds a predefined threshold:
\[
\Delta \text{pain} = \left| \text{pain}_{t} - \text{pain}_{t-1} 
\right| \geq \delta,
\]
where $\delta = 1$ corresponds to the smallest observable change 
on the NRS. Events are labeled according to the direction of 
change, either increase or decrease. This study focuses on 
pain-increase events, which are most relevant to acute pain 
escalation during procedures; pain-decrease events were excluded 
as a focused scope decision, though the framework generalizes 
naturally to bidirectional changes. We set $\delta=1$, the smallest 
discrete change on the NRS, which yields sufficient event density 
for rater-specific analysis. This threshold was selected to 
maximize event density while remaining clinically interpretable; 
inspection of an alternative threshold ($\delta=2$) did not 
qualitatively alter the observed rater-dependent patterns.

This rater-specific event definition captures rater decision points 
and preserves differences in how patients, nurses, and clinicians 
perceive and report changes in pain, without assuming clinical 
significance or causal intervention.

\subsection{Event-Aligned Physiological Window Extraction}

For each pain-change event, a fixed-length physiological window 
around the event time is extracted. Specifically, for an event 
occurring at time $t_0$, physiological signals are extracted within 
the interval $[t_0 - T_{\text{pre}}, t_0 + T_{\text{post}}]$, 
where $T_{\text{pre}} = 15$~s and $T_{\text{post}} = 30$~s. These 
window lengths were chosen to capture short-term autonomic 
responses --- consistent with known EDA and HR response latencies 
of 1--15~s following nociceptive stimuli --- preceding pain 
escalation without encroaching on neighboring events, while the 
post-event window allows observation of signal recovery. This 
framework is most appropriate for acute procedural pain settings; 
generalization to chronic pain, where relevant dynamics may evolve 
over longer timescales, remains an open direction for future work.

Physiological signals are standardized within each subject to 
reduce inter-individual baseline differences and to emphasize 
relative temporal dynamics. Events without sufficient physiological 
coverage within the analysis window are excluded to maintain 
consistency across subjects and raters. For each event and signal, 
we computed a pre-event slope feature as the least-squares linear 
trend (slope) of the within-subject standardized signal over 
$[-15,0]$~s, yielding one slope value per event and signal. 
Confidence bands in event-aligned figures represent $\pm$1 standard 
error of the mean, computed across subjects contributing at least 
one event for each rater, reflecting between-subject variability 
rather than within-subject repeated-measures uncertainty.

\subsection{Event-Centered Aggregation and Analysis}

Event-aligned physiological windows are aggregated according to 
rater group. For each physiological signal, the mean standardized 
response is calculated at each relative time point across events.

Formally, let $x^{(r)}_{i,s,e}(t)$ denote the standardized 
physiological signal $s$ for subject $i$ associated with event $e$ 
from rater $r$, aligned such that $t=0$ corresponds to the 
reported pain-change event. The event-centered average response 
for signal $s$ and rater $r$ is defined as follows:
\[
\bar{x}^{(r)}_s(t) = \frac{1}{N_r} \sum_{i,e} x^{(r)}_{i,s,e}(t),
\]
where the summation is over all events $e$ from subjects $i$ 
contributing to rater $r$, and $N_r$ denotes the total number of 
events for that rater.

This event-centered aggregation facilitates direct comparison of 
physiological dynamics preceding pain changes reported by different 
raters, without requiring a shared ground truth or combining 
rater-specific data.

\subsection{Inter-Rater Agreement and Signal-Level Comparisons}

To quantify disagreement in reported pain changes, we compare the 
magnitude of pain differences between rater pairs across temporally 
proximate assessment intervals within the same procedure, using all 
pairs of ratings recorded within a common time window of 
10~minutes. Where raters assessed pain at different times, the 
nearest available rating within this window was used; unpaired 
intervals were excluded. Distributions of absolute pain differences 
are summarized to characterize inter-rater disagreement.

The relationship between physiological signals and pain changes 
for each rater is examined by analyzing temporal patterns in 
event-aligned responses and by evaluating the relative importance 
of physiological features derived from pre-event windows. Feature 
importance is assessed using standardized linear models fitted 
independently for each rater group, allowing comparison of signal 
contributions without imposing a shared model across raters. These 
models are used for interpretability and comparison of signal 
contributions rather than for predictive performance. Given the 
modest event counts (41--47 per rater) and the repeated-measures 
structure of the data --- whereby multiple events derive from the 
same participant --- these results should be interpreted as 
exploratory rather than inferential. Future work should apply 
mixed-effects models to account for within-subject correlation 
across events.

Together, these analyses provide a rater-specific perspective on 
the association between wearable physiological signals and changes 
in reported pain.

\begin{figure}[t]
  \centering
  \includegraphics[width=\columnwidth]{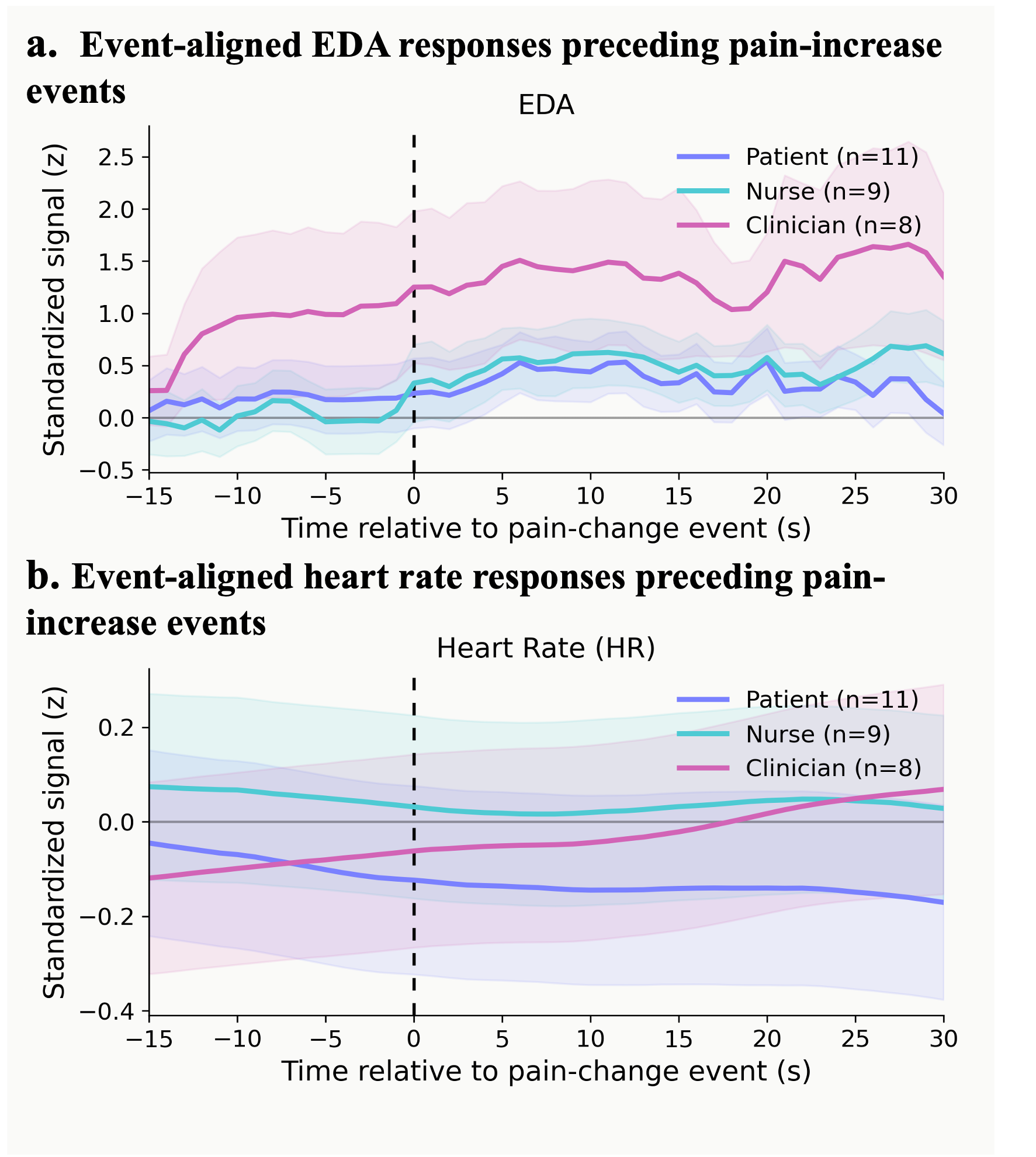}
  \caption{Event-aligned physiological responses preceding 
  pain-increase events.
  (a) Electrodermal activity (EDA).
  (b) Heart rate (HR).
  Shaded regions indicate $\pm$1 standard error of the mean across 
  subjects contributing at least one event for each rater.}
  \label{fig:eda_hr}
\end{figure}

\section{Results}
We first summarize rater disagreement and rater-dependent 
pain--physiology relationships using feature-level comparisons 
(Fig.~\ref{fig:results_overview}). We then examine event-aligned 
physiological time courses to illustrate the temporal dynamics 
underlying these feature differences (Fig.~\ref{fig:eda_hr}).

\subsection{Pain-Change Events Differ Across Raters}

The analysis first assessed whether pain-change events identified 
by different raters reflected a shared understanding of pain 
dynamics. Inter-rater disagreement was quantified as the absolute 
difference in reported pain changes between rater pairs 
(Fig.~\ref{fig:results_overview}(a)).

Substantial disagreement was observed across all rater pairs. The 
largest discrepancies occurred between patient- and 
clinician-reported pain changes, while patient--nurse and 
nurse--clinician pairs exhibited smaller but consistent 
differences. These findings suggest that pain-change events may 
not be uniformly perceived across raters, even when assessments 
occur within similar temporal windows.

Disagreement was not attributable to differences in reporting 
frequency. As shown in Table~\ref{tab:event_counts}, the median 
number of pain-change events per subject was comparable across 
raters, whereas the timing and identity of events differed. This 
pattern suggests that raters responded to different cues when 
identifying pain escalation, rather than simply reporting more or 
fewer changes.

Beyond disagreement in reported pain changes, we compared 
pre-event physiological trends using slope features extracted from 
the aligned windows. Fig.~\ref{fig:results_overview}(b) shows 
pre-event physiological slope features extracted from the 
$[-15,0]$~s window for each rater. The distributions differ across 
raters, suggesting that the direction and magnitude of pre-event 
autonomic trends may differ depending on who reports the pain 
change. In particular, clinician-reported pain increases tend to 
be preceded by stronger positive EDA and HR slopes compared with 
patient-reported events, while temperature slopes show weaker and 
less consistent separation.

\begin{table}[h]
\caption{Distribution of pain-change events by rater.}
\label{tab:event_counts}
\centering
\setlength{\tabcolsep}{3.5pt}
\renewcommand{\arraystretch}{1.15}
\begin{tabular}{|p{0.20\columnwidth}|p{0.22\columnwidth}|
p{0.18\columnwidth}|p{0.30\columnwidth}|}
\hline
\textbf{Rater} &
\textbf{Subjects with $\geq$1 event} &
\textbf{Total events} &
\textbf{Median events/subject (IQR)} \\
\hline
Patient & 12 & 47 & 4.0 (3.0--5.0) \\
\hline
Nurse & 12 & 46 & 4.0 (3.0--4.2) \\
\hline
Clinician & 12 & 41 & 4.0 (2.0--4.2) \\
\hline
\end{tabular}
\end{table}

\subsection{Rater-Dependent Physiological Dynamics Preceding 
Pain Changes}

The next analysis examined physiological dynamics aligned to 
rater-specific pain-change events using the proposed event-aligned 
framework. Fig.~\ref{fig:eda_hr} presents event-aligned EDA and 
HR, standardized within subjects and aligned to the reported 
pain-change time.

Exploratory rater-dependent physiological patterns were observed 
in the pre-event window. Clinician-reported pain increases were 
typically preceded by a gradual rise in EDA, whereas patient- and 
nurse-reported events exhibited smaller or more variable pre-event 
EDA changes. Heart rate dynamics further differentiated rater 
perspectives: patient-reported pain changes were preceded by a 
gradual decrease in HR, while clinician-reported events showed an 
increasing HR trend approaching the event. Nurse-reported events 
typically fell between these patterns. It should be noted that 
these patterns may partly reflect procedural confounds --- such as 
anticipation, movement, or medication timing --- that co-occur 
differently with each rater's assessment timing, rather than 
differences in pain perception per se.

These rater-specific differences emerged in the pre-event window, 
suggesting that physiology evolves differently before each rater's 
reported pain increase. To compare the relative contribution of 
each physiological feature across raters, we fit standardized 
linear models separately for each rater and report the resulting 
coefficients as an exploratory measure of relative signal 
association (Fig.~\ref{fig:results_overview}(c)); these models 
are used for interpretability only and not for predictive 
evaluation. EDA showed stronger associations with pain changes 
reported by nurses and clinicians than with patient-reported 
events, while heart rate contributed more prominently to 
clinician-reported pain changes. Skin temperature showed weaker 
and less consistent associations across raters, suggesting it 
may contribute limited discriminative value within the short 
pre-event windows examined here.

In summary, these exploratory results suggest that wearable 
physiological dynamics preceding reported pain changes may differ 
across rater groups. Given the modest event counts 
(41--47 per rater) and the repeated-measures structure of the 
data, these patterns should be interpreted cautiously and 
confirmed in larger studies with formal statistical testing.
\section{Discussion}
\label{sec:discussion}

This study examined pain assessment using a rater-aware, 
event-aligned approach and found exploratory evidence that 
relationships between wearable physiological signals and reported 
pain may depend on who reports the pain. Rather than treating 
inter-rater disagreement as annotation noise, the findings suggest 
that such disagreement may reflect systematic differences in 
perception, information sources, and decision criteria among 
patients, nurses, and clinicians.

\subsection{Rater-Dependent Pain Assessment and Physiological 
Alignment}

Substantial disagreement was observed across all rater groups in 
identifying pain-change events, even when assessments occurred 
within similar temporal windows. Disagreement was most pronounced 
between patient- and clinician-reported pain changes, while 
patient--nurse and nurse--clinician pairs exhibited smaller but 
consistent differences. These discrepancies appeared attributable 
to differences in how pain changes were interpreted, rather than 
to variation in reporting frequency.

Event-aligned analysis further suggested that physiological 
dynamics preceding reported pain changes may vary across raters. 
Clinician-reported pain increases were preceded by stronger and 
more consistent autonomic signatures, particularly in 
electrodermal activity, whereas patient- and nurse-reported events 
exhibited weaker or more variable pre-event responses. Heart rate 
dynamics also differed among raters, potentially reflecting 
distinct sensitivities to autonomic, behavioral, and contextual 
factors. It should be noted, however, that observed physiological 
patterns may partly reflect procedural confounds such as 
anticipation, movement, or medication timing, which differ in how 
they relate to each rater's assessment process. Future work should 
incorporate procedure-stage annotations to partially address these 
alternative explanations.

\subsection{Implications for Wearable-Based Pain Monitoring}

These results have potential implications for the design of 
wearable-based pain monitoring systems. Many existing approaches 
implicitly assume a single ground-truth pain label and optimize 
models accordingly. The present findings suggest that such 
assumptions may conflate distinct physiological processes and 
obscure meaningful differences in how pain is perceived and 
evaluated across clinical roles.

Adopting a rater-aware, event-centered perspective may more 
accurately reflect real-world clinical assessment practices. 
Rather than predicting a single pain score, future systems may 
benefit from modeling multiple rater perspectives or explicitly 
incorporating rater-specific decision criteria. One concrete 
direction is clinical decision support that flags discrepancies 
between patient self-report and clinician observation in 
real time, enabling more targeted intervention rather than 
enforcing agreement through label aggregation. Such a system 
would leverage the rater-aware framework presented here as a 
foundation for rater-specific signal interpretation.

\subsection{Limitations and Future Directions}

This study has several limitations. The cohort size was modest 
(n=24) and drawn from a single clinical context, which limits 
generalizability. The modest event counts (41--47 per rater, from 
12 subjects each) and the repeated-measures structure of the data 
mean that the reported patterns should be treated as exploratory; 
future work should apply mixed-effects models to account for 
within-subject correlation across events. The analysis focused 
exclusively on pain-increase events and did not examine 
pain-decrease dynamics, which may follow different physiological 
patterns. The use of two wearable platforms (Empatica E4 and 
EmbracePlus) introduces potential signal-level heterogeneity that 
future studies should address by using a single standardized 
device or explicitly assessing device effects. In addition, 
observed rater-dependent physiological patterns may reflect 
procedural confounds such as anticipation, movement, or medication 
timing, rather than pain perception per se. Skin temperature 
showed weak and inconsistent associations across all rater groups, 
suggesting it may contribute limited discriminative value within 
the short pre-event windows examined here; its utility over longer 
timescales warrants further investigation. Finally, the 
physiological signals analyzed were limited to those commonly 
available in wrist-worn devices; incorporating additional 
modalities such as phasic EDA components or HRV-derived features 
may provide complementary insights.

Future work will extend this framework to larger and more diverse 
clinical populations, explore adaptive and personalized event 
definitions, and integrate measures of rater confidence and 
uncertainty. Beyond pain assessment, rater-aware, event-aligned 
analysis may provide a generalizable approach for studying 
subjective clinical assessments in conjunction with continuous 
physiological data.

\section{Conclusion}

This study introduced a rater-aware, event-aligned framework for 
analyzing relationships between subjective pain assessments and 
continuous wearable physiological signals. By transforming sparse, 
rater-specific pain ratings into discrete pain-change events and 
aligning multimodal physiological data to these events, the 
proposed approach moves beyond single-label assumptions and 
explicitly incorporates rater-dependent pain assessment.

The results revealed substantial disagreement among patients, 
nurses, and clinicians in identifying pain changes, as well as 
exploratory evidence of rater-dependent physiological differences 
preceding reported pain increases. These findings suggest that 
pain--physiology relationships may not be rater-invariant, and 
that treating pain assessments as interchangeable labels may mask 
rater-specific physiological patterns.

By preserving temporal structure and rater identity, the proposed 
event-aligned analysis provides a principled foundation for 
integrating wearable physiological data with real-world clinical 
pain assessments. This rater-aware perspective may support the 
development of more interpretable, context-aware pain monitoring 
systems and inform future wearable-based clinical decision support 
tools.

\bibliographystyle{IEEEtran}
\bibliography{references}
\end{document}